\documentstyle[twoside,11pt,draft]{article}
\def\p-{$\pi^-\,$}

\begin{document}
\input epsf
%
\Large
\normalsize
\begin{flushright}
\today \\
ISN Grenoble report 00/48
\end{flushright}
\par
\vspace{1cm}
\begin{center}
{\Large \bf Origin of the high energy proton component below the geomagnetic cutoff 
in near earth orbit}
\vspace{0.3cm}
\par
    L. Derome, M. Bu\'enerd, A. Barrau, A. Bouchet, A. Menchaca-Rocha \footnote{Permanent 
address: Instituto de Fisica, IFUNAM, Mexico}, and T. Thuillier
\par
\vspace{0.2cm}
{\sl Institut des Sciences Nucl\'{e}aires, IN2P3, 
53 av. des Martyrs, 38026 Grenoble cedex, France}
\par
\normalsize
\end{center}
\vspace{0.5cm}
\begin{center}
\parbox{16cm}{\underline{Abstract}: The high flux proton component observed by AMS below 
the geomagnetic cutoff can be well accounted for by assuming these particles to 
be secondaries originating from the interaction of cosmic ray protons with the atmosphere.
Simulation results are reported. }
\end{center}
\setcounter{page}{1}
%
The existence of a high flux proton component below the earth geomagnetic cutoff (GC), was 
reported recently by the AMS collaboration \cite{PROT}. The observed spectrum has an 
intensity about equal to the cosmic primary proton flux above GC. It has a maximum 
at low momenta and extends from the momentum threshold of the spectrometer at 
$\approx$0.4~GeV/c, up to the GC around 10~GeV/c in the equatorial region, where the high 
momentum tail merges with the primary proton spectrum above GC.
\par
It has been demonstrated long ago that such subGC component cannot be part of the primary 
proton flux \cite{STORM,SI60}. It has to be a secondary product of the primary cosmic ray 
(CR) flux on earth. This note reports on an attempt both to identify which part of the 
incident CR flux on earth generates this spectrum and by which mechanism, and to investigate 
its dynamical status with respect to the trapping process observed for other circum-terrestrial 
populations of particles. 
\par
A similar phenomenon was observed previously with the very high intensity low energy proton 
flux measured in the Van Allen belts, at kinetic energies typically below 100~MeV and for 
distances typically beyond 1.2 earth radius, which was interpreted as originating from 
secondary neutrons produced on atmospheric nuclei by CR protons \cite{NP}. Many low energy 
neutrons are produced by nuclear evaporation, or spallation, or preequilibrium nuclear 
emission, which can constitute the source term driving the equilibrium of this population.
\par
The situation is quite different for the population of higher momentum protons, beyond 
0.5~GeV/c, observed here. It is too high in energy to be accounted for by the above quoted 
mechanism, neither could it be by neutrons from the direct charge exchange reaction 
$pn\rightarrow np$, induced by high energy CR protons of nuclei, because of highly 
unfavored kinematics and small cross sections of the diffractive production mechanism, and 
very small probability of neutron decay below the altitude of AMS. 
\par 
The two most significant sources of secondary protons over the observed energy range are the 
$p+{\cal A}$ and $^4$He$+{\cal A}$ collisions, {$\cal A$} standing for atmospheric nuclei, the 
CR flux of higher nuclear masses being too small to contribute significantly.
\par
Considering primary protons as a likely dominant source of the observed subGC protons, two 
extreme situations are met over the explored range of latitudes: In the equatorial region, 
primary and secondary spectra do not overlap, incident protons are above GC, then with 
E$\ge$10~GeV, and produce a spectrum of secondaries (below GC) with E$\le$10~GeV, whereas 
in the polar region (no GSC) primary and secondary protons are totally mixed in a spectrum 
extending down to the 0.1~GeV kinetic energy range. This latter feature implies that a 
variety of reaction mechanisms, from nuclear in the low energy range to subnucleonic 
hadronic in the upper range, could contribute to generate the observed subGC population. 
\par
Experimentally, the $p{\cal A}\rightarrow pX$ reaction on light nuclei has been studied 
in several experiments over a broad range of incident energies \cite{ALLREF,BA79,GE80,BA84,
BO87,AB92}, (see \cite{PP2PX,RO75} for inclusive $pp\rightarrow pX$ studies) providing a 
body of data rich enough for systematic trends to be identified and for a consistent 
physical picture of the dominant mechanisms to emerge.  
Two components were observed in the measured proton spectra, which could be qualitatively
identified to two different regimes of collision. One is forward peaked, dominated with 
leading particle effects, and corresponds to peripheral collisions and low orders of 
multiple intranuclear proton-nucleon collisions. The closeness of the experimental forward
proton multiplicites measured in $p{\cal A}$ \cite{AR78} and $pp$ \cite{RO75} collisions 
support this picture and shows in addition that protons measured in the first reaction 
are dominantly hadronic products of $pp$ collisions. This is also consistent with the known 
proton mean free path in nuclei \cite{TA81}. The other component corresponds to more central 
collisions and higher order type of multiple intranuclear collisions. In the following, these 
two components will be referred to as {\sl quasi elastic} (QE) and  {\sl deep inelastic} 
(DI) respectively, following ref~\cite{BA84}. 
In ref~\cite{BA84}, the first component was observed for low energy protons only in the 
angular range allowed by quasi free scattering kinematics, whereas the other component was 
observed also at backward angles in the region of zero rapidity, kinematically not 
accessible to free proton-nucleon processes. 
The two components have been found to exhibit significantly different dependence on the 
nuclear target mass, supporting the above description \cite{GE80,BA84}.  
\par
In this context, the following features of the reactions relevant to the present study, 
have been observed: 
a) The invariant cross section for proton production versus transverse kinetic energy, can 
be described with a simple exponential over the transverse mass, with a slope weakly 
dependent on the rapidity \cite{AB92}. 
b) At backward angles, 
a limiting production mechanism has been observed for incident energies above 
8.5~GeV/c \cite{BA79,BA84,BO87}, the cross-section showing little incident energy 
dependence, with the onset of the limiting behaviour occuring as low as 1-2~GeV 
kinetic energy per nucleon for protons produced within the range 0.4-1~GeV/c \cite{GE80}. 
It was shown in ref~\cite{BA84} that the target-like proton production can be described 
with a simple parametrization over the full angular range (see ref~\cite{DA90} and the 
quoted refs in this paper for the theoretical aspects). 
\par
The inclusive spectrum of protons at the altitude of AMS (390-400km) has been calculated 
by means of a computer simulation program built to this purpose. 
CR particles are generated with their natural abundance and momentum distributions 
\cite{ABOND} corrected for the solar modulation effect \cite{SOLMOD}, at a distance of 5 
terrestrial radii, or 2 St\"ormer length units \cite{STORM} in order to ensure the 
generation point to be always in appropriate initial conditions for the earth magnetic 
field. They are propagated inside the earth magnetic field \cite{CHP}, using 4th order 
adaptative Runge-Kutta integration of the equation of motion. Particles are allowed to 
interact with atmospheric nuclei (mainly $^{14}$N and $^{16}$O, see ref~\cite{ATM} for the 
model of atmosphere) and produce secondary protons with cross sections and multiplicities as 
discussed below. Each secondary proton is then propagated and allowed to collide as in the 
previous step. 
A reaction cascade can thus develop through the atmosphere. The reaction products are 
counted when they cross the virtual sphere at the altitude of the AMS spectrometer, upward 
and downward. Particles undergo energy loss by ionisation before and after the interaction. 
Multiple scattering effects have not been included at this stage. Each event is 
propagated until the particle disappears by either colliding with a nucleus, or being 
stopped in the atmosphere, or escaping to outer space beyond twice the production altitude. 
A 2000~s time out protection in the program has never been called for the simulated 
sample. It must be noted that particles are counted each time they cross the sphere of 
detection altitude. The contributions of trapped particles are thus weighted statistically 
with their numbers of crossings, which increases in proportion their contribution to the  
final spectrum. Some empirical cuts over the initial particle kinematics allowed significant 
computer time saving by rejecting kinematically irrelevant events. 
\par
\begin{figure}[htb]
\begin{center}
\vspace{-1cm}
\hspace{-2cm}
\epsfysize=10cm
\epsfbox{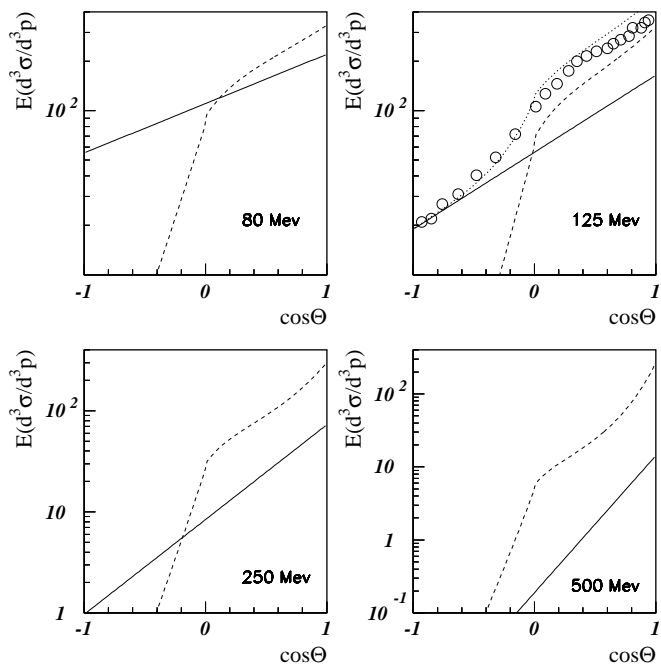} \\
\end{center}
\vspace{-0.5cm}
\caption{\small\it Angular distributions of the two proton components from $pp$ collisions 
included in the simulation, for low secondary proton energies: Parametrization of 
ref~\cite{AB92} data (dashed line), and parametrization from ref~\cite{BA84} (solid line). 
In the upper right plot, the sum of the two components is compared to the data from 
\cite{BA84}.}
\label{PA2PX}
\end{figure}

The secondary nucleon spectrum generated has to cover two orders of magnitudes in kinetic 
energy, 
between about 100~MeV and 10~GeV. The main component of proton production cross section was 
obtained by means of analytical relations fitted to the 14.6~GeV $p+Be$ data \cite{AB92} 
using the slope parameters given in this reference for the transverse mass distribution, 
whereas the slowly varying rapidity distributions were fitted with a polynomial. The 
scaling properties have been checked with the FRITIOF/PYTHIA (L\"und) event generator 
\cite{FRIT}. Since this generator is not expected to account for the very low energy and 
backward proton emission (target-like to negative rapidities), this latter component was 
incorporated using of the parametrization given in ref~\cite{BA84}. The respective 
contributions to the total multiplicity-weighted proton production cross-section, were 
352~mb for the QE component and 88~mb for the DI components. Cross sections on atmospheric 
nuclei were renormalized from the original data or parametrizations obtained on different 
nuclei, using ratios of geometrical cross sections. Figure~\ref{PA2PX} shows the two proton 
components as a dashed line (QE), and solid line (DI) from $p+C$ at 7.5~GeV. For 125~MeV 
protons, it is seen that the data \cite{BA84} are pretty well reproduced by the sum of the 
two components, providing sound grounds to the present calculations. For each event, the 
proton multiplicity was generated using a Poisson distribution with the experimental mean 
value $\mu$=1.7 from ref~\cite{AR78}. Neutron spectra and multiplicities were taken 
identical to proton's.
\par
%
\begin{figure}[htb]
\begin{center}
\vspace{-1cm}
\hspace{-2cm}
\epsfysize=10cm
\epsfbox{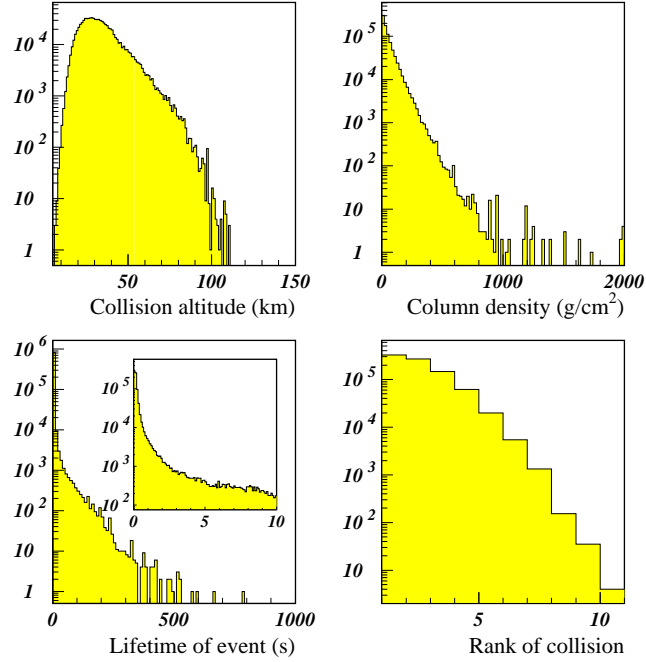} \\
\end{center}
\vspace{-0.5cm}
\caption{\small\it Distributions of production altitude (km), thickness of matter crossed 
(g/cm$^2$), propagation time (s) between particle production and detection, and 
distribution of the rank of the collision in the atmospheric cascade for a given event, 
for the simulated sample of events.}
\label{GENE}
\end{figure}

A sample of 6 10$^7$ particles was generated over the terrestrial sphere (120 hours of CPU 
computing time), of which 2.7 10$^6$ reached the atmosphere, corresponding to a sampling 
time of the incident flux of the order of 0.5~ps. A cut requiring the accepted events to be 
inside the angular acceptance of AMS \cite{PROT} was applied to the detected protons.
Figure~\ref{GENE} shows some basic features of the simulated sample. The four histograms 
show respectively, the distribution of the altitude of the interaction points (upper 
left), the thickness of matter crossed by particles (upper right), the lifetime of 
particles between production and detection (lower left), and the rank of the event in the 
atmospheric collision cascade (lower right). 
%
%
The distribution of the proton production altitude has a maximum around 25~km (mean value 
39~km), and then decays approximately exponentially, following the atmospheric density 
profile. The thickness of matter crossed by the particles is distributed over a range 
extending up to about 600~g/cm$^2$, i.e., about 10 collision lengths, as it could be 
expected from the known range of the incident particles. 
The lifetime of the particles extends up to around 600~s. The insert shows the distribution 
between 0 and 10~s. The lifetime vs momentum correlation obtained (not shown) is in 
qualitative agreement with the backtracing calculations reported in \cite{PROT}.
\par
It is seen on the lower right histogram that more than 50\% of the detected protons are not 
produced in the first collision, and originate from up to the 10th collision generation in 
the atmospheric cascade for a given incident proton. An expected correlation is observed 
between collision rank and production altitude.
\par
%
\begin{figure}[hp]
\begin{center}
\vspace{-1cm}
\hspace{-2cm}
\epsfysize=9cm
\epsfbox{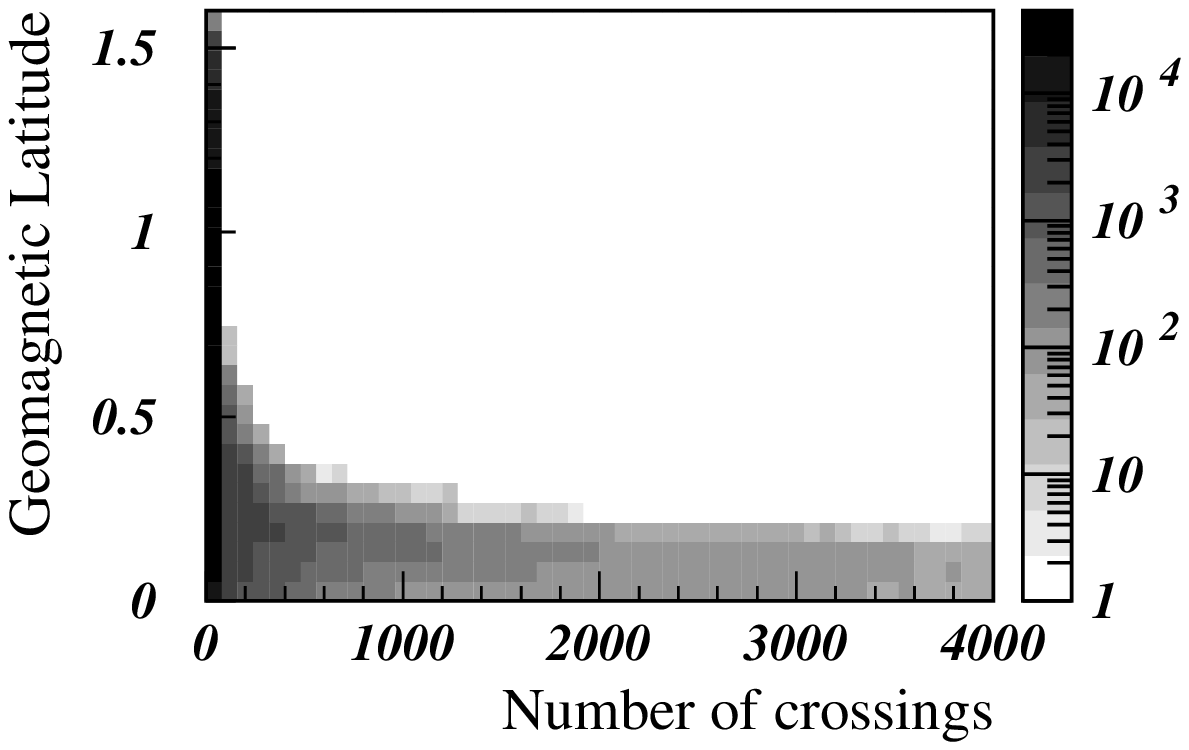} \\
\end{center}
\vspace{-0.5cm}
\caption{\small\it Two dimensionnal plot of the distribution of the CGR (geomagnetic) 
latitude \cite{CGR} versus number of crossings of the detection altitude, showing that a 
population of particles is trapped in the equatorial region.}
\label{TRAP}
%
%
\begin{center}
\hspace{-2cm}
\epsfysize=9cm
\epsfbox{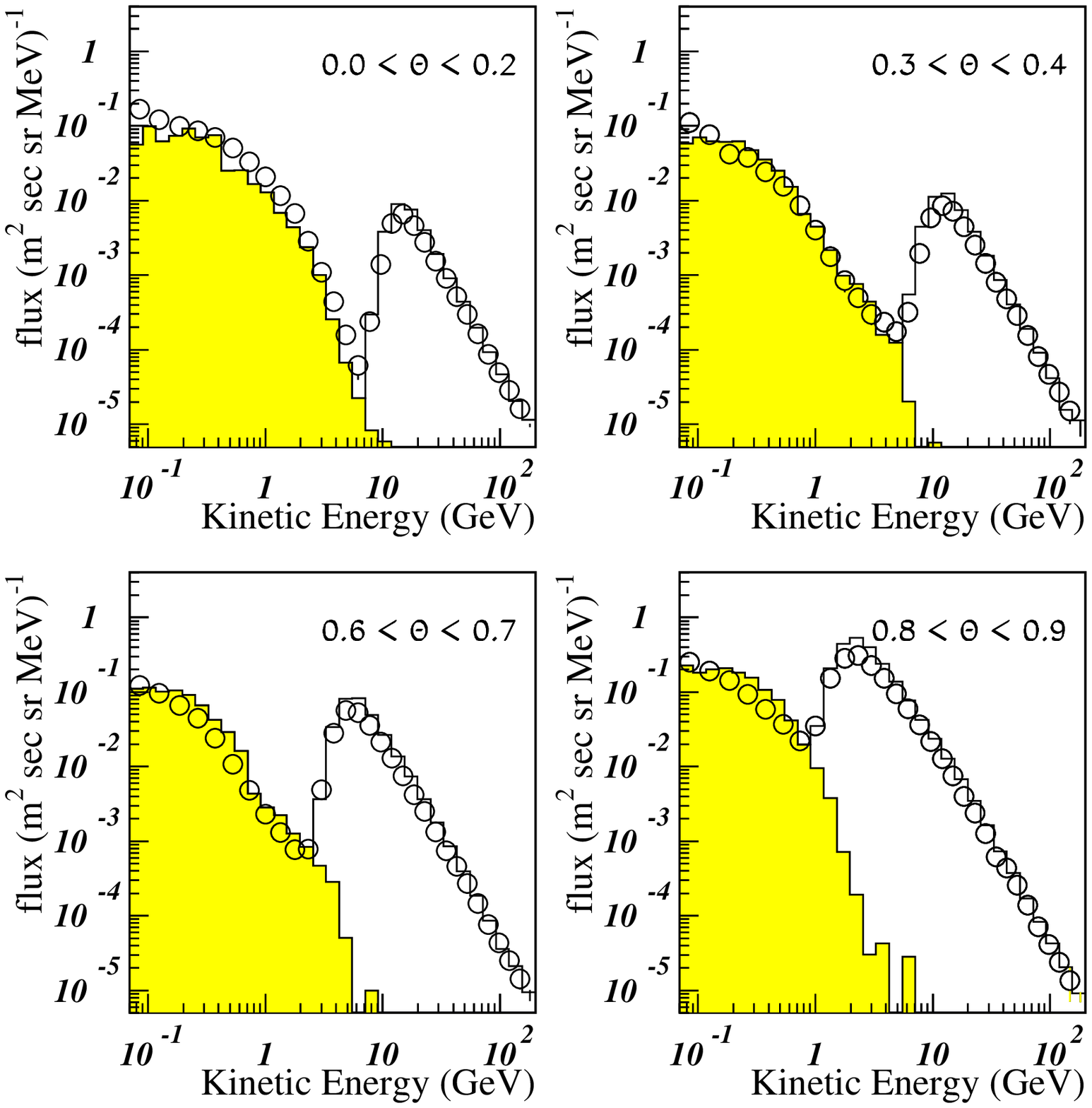} 
\hspace{0.2cm}
\epsfysize=9cm 
\epsfbox{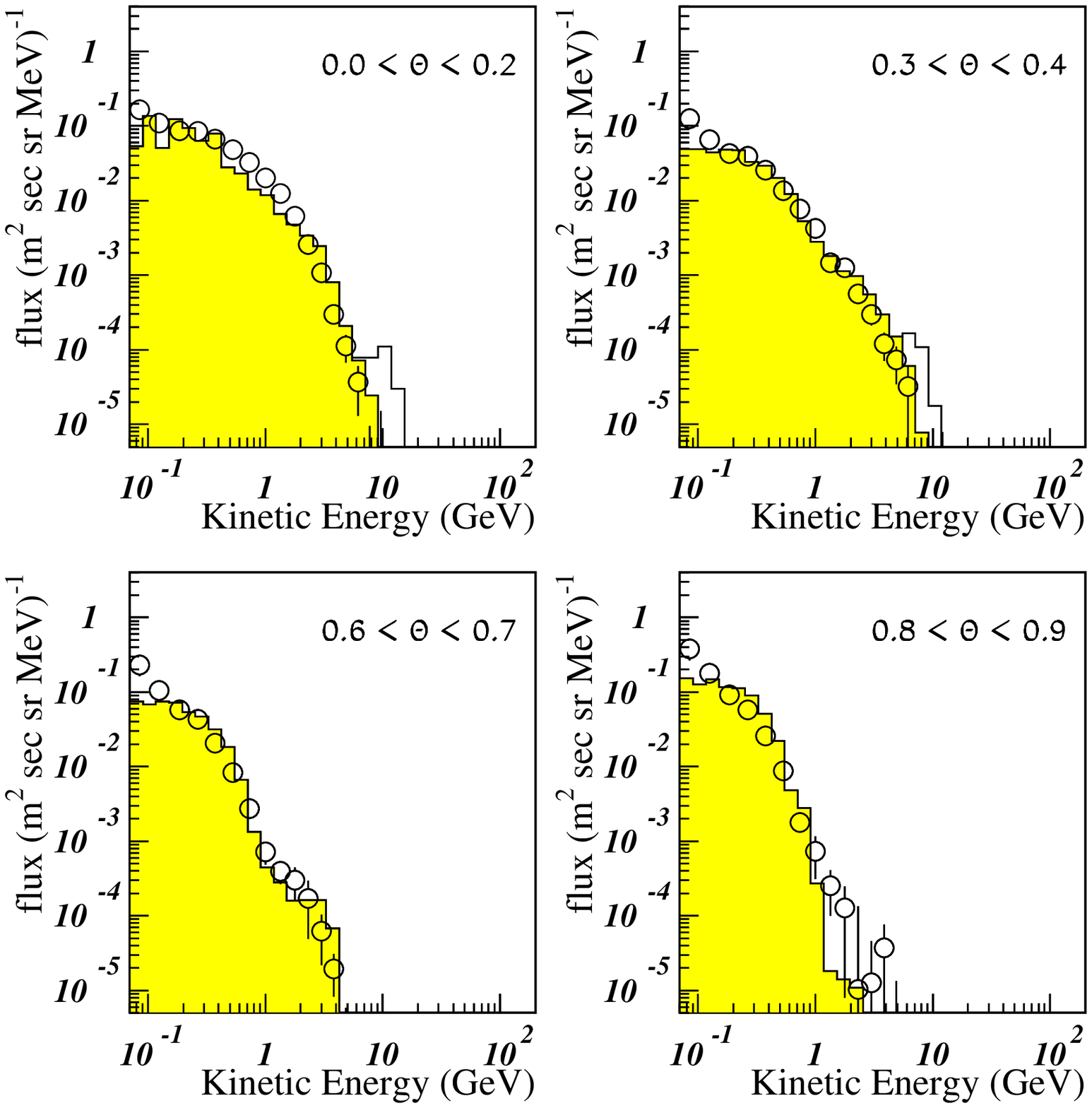} \\
\end{center}
\caption{\small\it Experimental kinetic energy distributions from \cite{PROT} for a sample
of bins in latitudes (open circles), compared to the results of the simulation (full line)
described in the text, for downward (left) and upward (right) protons.}
\label{DISTRIB}
\end{figure}
Figure~\ref{TRAP} displays an important feature of these results with the scatter plot of
the detection latitude versus the number of crossings of the detection altitude. 
It shows that events with large crossing multiplicities are strongly correlated with
low latitudes, thereby showing that a population of trapped particles confined in the 
equatorial region $\approx\pm$0.3~rad is expected to be observed experimentally. 
\par
In addition, the following other features are observed:\\
- A population of longlived particles, with flight times up to several 10$^2$s, is found in 
the polar region, more than 95\% of them having only one or two crossings, i.e., having 
very long flight path between two crossings. The energy distribution of these particles is 
the same as that of particles at lower latitudes. Their trajectories then make large loops 
away from earth. \\
- The angular distribution of the detected protons with respect to the incident proton 
direction, resulting from the folding of the (forward peaked) cross section of the proton 
production reaction with the (outward peaked) momentum acceptance of the magnetospheric 
transport system, is found to be approximately evenly distributed over the forward 
hemisphere and decaying rapidly at backward angles. 
\par
Figure~\ref{DISTRIB} shows the experimental kinetic energy distributions of downward 
(left) and upward (right) protons measured for some bins of latitude, compared to the 
results of the simulation. No free parameter is used for normalization to the data:  
The calculated results are entirely determined by the physics input to the calculation. 
It is seen that the agreement between the data and the simulation result is remarkably
good, at all latitudes and for both the inward and outward flux. In particular, the cutoff 
region is particularly well reproduced, which indicates that the processing of the particle 
dynamics and kinematics is good. The shaded histograms in the figure correspond to secondary 
particles in the simulation. The fraction of events originating from the DI component of 
the proton production cross section described previously, vary from about 10\% in the 
equatorial region up to 25\% in the polar region, with a momentum distribution peaking 
at low kinetic energy and distributed below 500~MeV.
\par
It must also be noted that the broad enhancement observed experimentally in the 1~GeV 
region in the subGC spectrum for the equatorial latitude compared to the other latitudes, 
is quite well reproduced. It has been verified that this enhancement is mainly due to 
the population of trapped particles above 10 crossings as shown on figure~\ref{TRAP}. The 
energy distribution of this population is confined below about 3~GeV since particles 
circling vertically in the earth magnetic field around the mean AMS altitude (380~km) have 
a Larmor radius of approximately 110~km per GeV/c. 
\par
The calculated flux at very low momenta shows a more or less significant deficit compared 
to the data for most latitudes. Since more than 50\% of the detected particles originate 
from secondary nucleon interaction with the atmosphere, the concerned incident energy range 
extends down to values where contributions of nuclear processes not taken into account by 
the event generator, could account for the observed deficit. A defect of the particle 
trapping processing at low momenta is also possible however. 
\par
The $^4$He cosmic flux should also contribute to the studied population. This 
contribution should roughly scale with the ratio of incoming flux of $^4$He and protons, 
since after the primary interaction, the atmospheric cascade should develop the same way. 
It is therefore expected to be about one order of magnitude smaller than the cosmic proton 
contribution \cite{ABOND}, which would not change significantly the results. Detailed 
investigations are in progress on the above points.
\par
The AMS measurements constitute a new input to the calculations of atmospheric neutrino 
production (see refs \cite{GA00} for example). New calculations are clearly required to 
evaluate the effect of the observed subGC component of the proton flux on the 
neutrino flux. It must be observed that the two proton populations, above and 
below GC, do not have the same interaction probability and angular distribution, and then 
will not affect equally the atmospheric neutrino flux.    
\par
In summary it has been shown that the high intensity flux of protons observed below the 
geomagnetic cutoff by the AMS experiment can be well reproduced by assuming that this flux 
originates from the interaction of the primary proton Cosmic Ray flux with the atmosphere. 
The technical means developed in this work are being applied to the investigation of the 
other particle populations observed by AMS. They are also potentially useful for similar 
calculations concerning other experiments such as 
the atmospheric antiproton component in satellite and balloon experiments, as well as for 
many other related processes and particle populations.
Detailed results will be reported in a more comprehensive paper in preparation. 
\par 
\end{document}